\documentclass{svproc}
\usepackage{slashed}
\usepackage{graphicx}
\usepackage{url}

\begin{document}
\mainmatter              
\title{Meson-baryon Scattering in Extended-on-mass-shell scheme at $\mathcal{O}(p^3)$}
\titlerunning{Meson-baryon scattering}  
%
\author{Junxu Lu\inst{1,2} \and Lisheng Geng\inst{1} \and Xiulei Ren\inst{3} \and Menglin Du\inst{4} }
\authorrunning{Junxu Lu et al.} 
%
\tocauthor{Junxu Lu, Lisheng Geng, Xiulei Ren, and Menglin Du}
\institute{School of Physics and
Nuclear Energy Engineering \& International Research Center for Nuclei and Particles in the Cosmos \&
Beijing Key Laboratory of Advanced Nuclear Materials and Physics,  Beihang University, Beijing 100191, China\\
\email{lisheng.geng@buaa.edu.cn}
\and
Groupe de Physique Th\'{e}orique, IPN (UMR8608), Universit\'{e} Paris-Sud 11, Orsay, France\\
\and
Institut f\"{u}r Theoretische Physik II, Ruhr-Universit\"{a}t Bochum, D-44780 Bochum, Germany\\
\and
Helmholtz-Institut f\"ur Strahlen- und Kernphysik and Bethe Center for Theoretical Physics, Universit\"at Bonn, D-53115 Bonn, Germany}

\maketitle              

\begin{abstract}
In this present work, we study the scattering of a pseudoscalar meson off one ground state octet baryon in covariant baryon chiral perturbation theory up to the next-to-next-to-leading order.
We remove the power counting breaking terms with the extended-on-mass-shell scheme. We perform the first combined study of the pion-nucleon and kaon-nucleon scattering data and show that the covariant baryon Chiral perturbation theory can provide a reasonable description of the experimental data for both channels.
\keywords{meson-baryon scattering, extend-on-mass-shell scheme, covariant BChPT}
\end{abstract}
\section{Introduction}
Chiral perturbation theory (ChPT), as a low-energy effective field theory of QCD, plays an important role in our understanding of the non-perturbative strong interactions~\cite{Weinberg:1978kz}. Elastic meson-baryon scattering is a fundamental process that not only can test our understanding of the strong interaction but also plays a relevant role in the studies of the
properties of single and multi baryons. Because of the large nonzero baryon masses $m_0$ in the chiral limit, lower order analytical terms appear in nominal higher order loop calculations, and therefore a consistent power counting is lost~\cite{Gasser:1987rb}. In the past three decades, several solutions have been proposed. The most studied ones are the heavy baryon ChPT~\cite{Jenkins:1990jv,Bernard:1995dp}, the infrared (IR) baryon ChPT~\cite{Becher:1999he}, and the extended-on-mass-shell (EOMS) BChPT~\cite{Gegelia:1999gf,Fuchs:2003qc}. Among them, the EOMS BChPT turns out to be more appealing because it satisfies all the symmetry and analyticity constraints and converges relatively faster~\cite{Geng:2008mf,Geng:2013xn}.

In this talk, we present our work on meson-baryon scattering up to next-to-next-to-leading order in the EOMS scheme.

\section{Theoretical Formalism}
In the isospin limit, the standard decomposition of the meson-baryon scattering amplitude reads~\cite{Gasser:1987rb},
\begin{equation}\label{stde1}
  T_{MB}=\overline{u}(p',s')\left[D+\frac{i}{m_i+m_f}\sigma^{\mu\nu}q'_{\mu}q_{\nu}B\right]u(p,s),
\end{equation}
where the $p(p')$ and $q(q')$  are the momentum for initial (final) baryons and mesons, respectively.

In order to calculate the meson-baryon scattering amplitudes up to the leading one-loop order,  i.e., $\mathcal{O}(p^3)$, we need the following meson-meson and meson-baryon Lagrangians:
\begin{equation}
\mathcal{L}_\mathrm{eff}=\mathcal{L}^{(2)}_{MM}+\mathcal{L}^{(4)}_{MM} + \mathcal{L}^{(1)}_{MB}+\mathcal{L}^{(2)}_{MB}+\mathcal{L}^{(3)}_{MB},
\end{equation}
where the explicit form of each term can be found in Ref.~\cite{Frink:2006hx,Oller:2007qd,Oller:2006yh}.

\begin{table}\label{final-LEC}
\centering
\caption{Independent (combinations of) LECs  contributing to $\pi N$ and $KN$ scattering. Note that in the fitting process, we have neglected the contributions from tree level at $\mathcal{O}(p^3)$ in the $KN$ channels.}
\begin{tabular}{c|c|c|c|c|c|c}
\hline\hline
$\pi N$ & FIT &  $KN_{I=0}$ & FIT &  $KN_{I=1}$ & FIT  \\
\hline
 $b_1+b_2+b_3+2b_4$&$-7.64(6)$& $b_3-b_4$       &$-0.767(1)$& $b_1+b_2+b_4$  &$-0.419(2)$ &$[\mathrm{GeV^{-1}}]$\\
 $b_5+b_6+b_7+b_8$ &$1.42(2)$&  $b_6-b_8$       &$0.126(1)$&  $2b_5+2b_7+b_8$&$0.429(2)$  &$[\mathrm{GeV^{-2}}]$\\
 $c_1+c_2$         &$1.34(1)$&  $4c_1+c_3$      &$0.604(3)$&  $4c_2+c_3$ &$0.616(1)$ &$[\mathrm{GeV^{-1}}]$\\
 $2b_0+b_D+b_F$    &$-1.36(6)$& $b_0-b_F$       &$0.093(1)$&  $b_0+b_D$ &$-0.090(3)$ &$[\mathrm{GeV^{-1}}]$\\
 $d_2$             &$0.61(2)$&  $d_1+d_2+d_3$   &0  &         $d_1-d_2-d_3$&0  &$[\mathrm{GeV^{-4}}]$\\
 $d_4$             &$3.25(6)$&  $d_4+d_5+d_6$   &0  &         $d_4-d_5+d_6$ &0 &$[\mathrm{GeV^{-2}}]$\\
 $d_8+d_{10}$      &$1.45(3)$&  $d_7-d_8+d_{10}$&0  &         $d_7+d_8+d_{10}$ & 0&$[\mathrm{GeV^{-3}}]$\\
 $d_{49}$          &$-0.32(12)$&$d_{48}+d_{49}+d_{50}$&0  &  $d_{48}+d_{49}-d_{50}$ &0 &$[\mathrm{GeV^{-2}}]$\\
\hline
 $\chi^2/d.o.f$             &$0.154$& $\chi^2/d.o.f$ &0.971& $\chi^2/d.o.f$ & 0.471 \\
\hline
\end{tabular}
\end{table}

It should be noted that not all of the remaining $\mathcal{O}(p^2)$ and $\mathcal{O}(p^3)$ terms contribute to a specific process. Particularly, for pion-nucleon and kaon-nucleon scattering, only 24 out of the total 37 LECs contribute.  They are tabulated in Table.~\ref{final-LEC}.

Because the baryon mass at chiral limit does not vanish, the power counting rules are violated~\cite{Gasser:1987rb}. Within the EOMS scheme, as shown in Ref.~\cite{Fuchs:2003qc}, these power counting breaking terms are all analytic and can be absorbed into low energy constants. Thus in the present work, we apply the $\overline{MS}-1$ first to absorb the ultraviolet divergence and then the EOMS scheme to remove the PCB terms to restore the power counting rules.

\section{Fitting and Results}
In the present work we focus on the $\pi N$ and $K N$ channels, because only for these channels partial wave phase shifts are available.

\begin{figure}
\centering
\begin{tabular}{ccc}
{\includegraphics[width=0.25\textwidth]{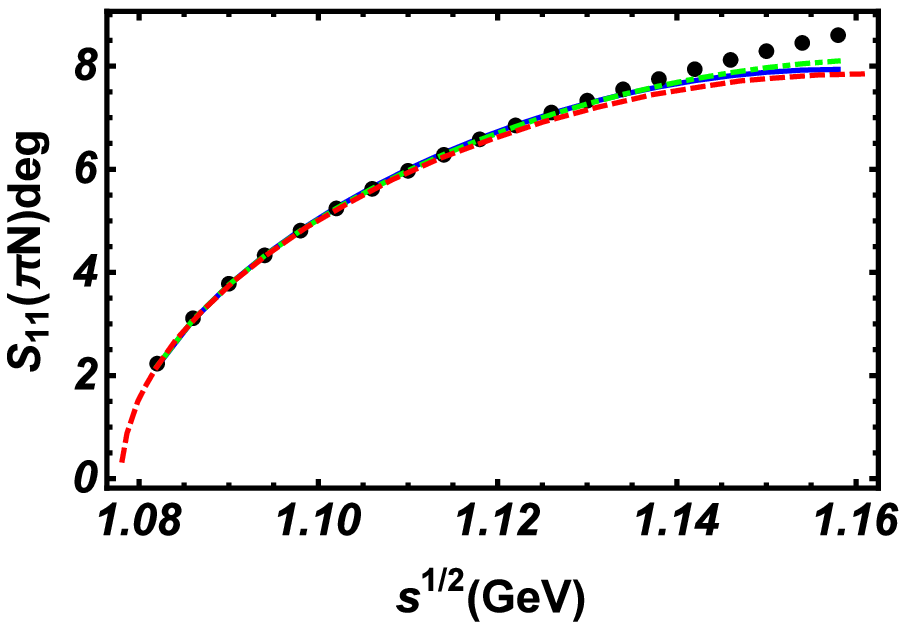}}&{\includegraphics[width=0.25\textwidth]{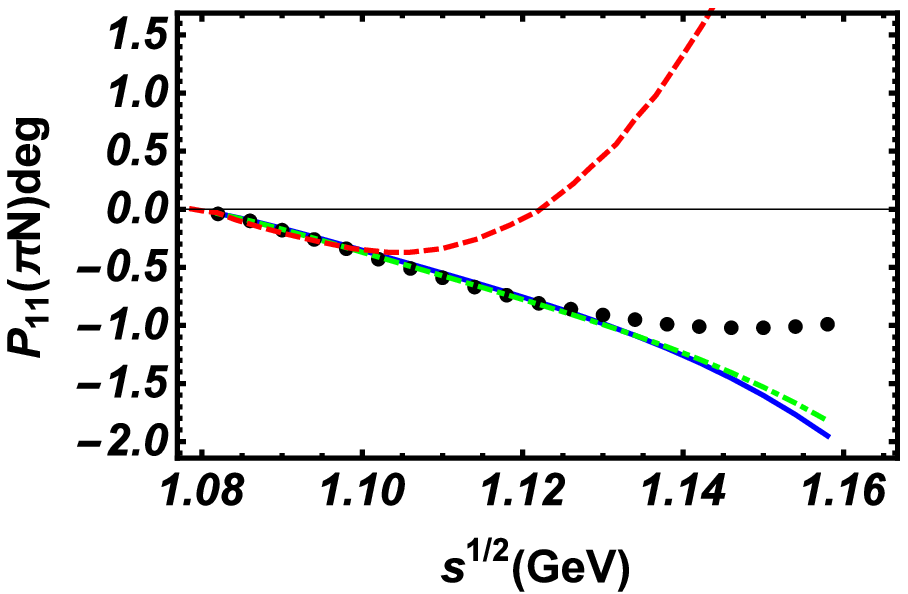}}& {\includegraphics[width=0.25\textwidth]{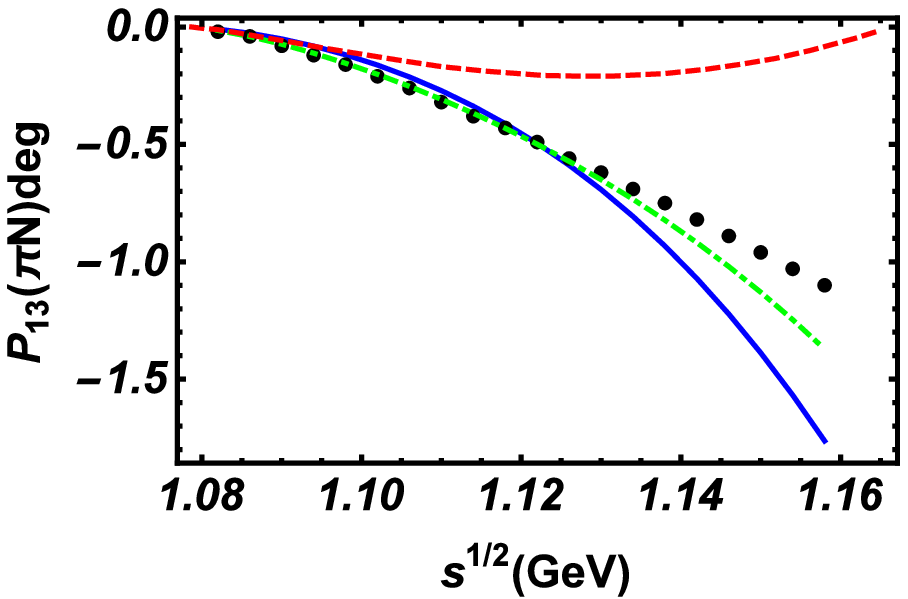}} \\
{\includegraphics[width=0.25\textwidth]{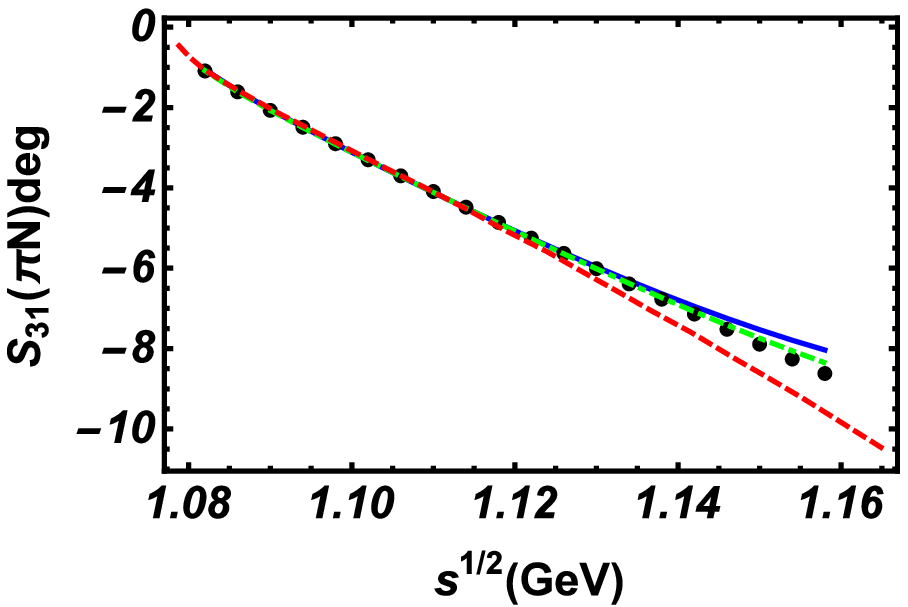}}&{\includegraphics[width=0.25\textwidth]{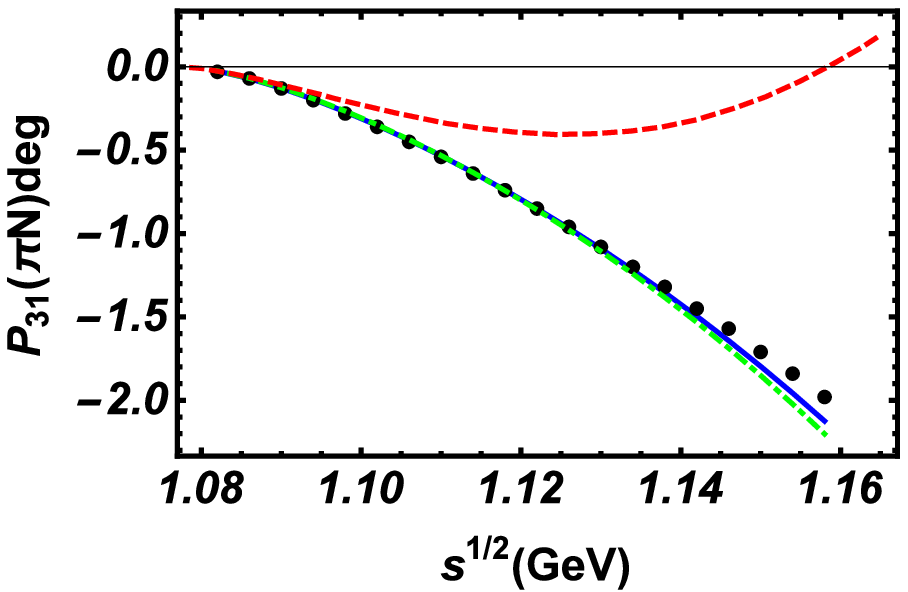}}& {\includegraphics[width=0.25\textwidth]{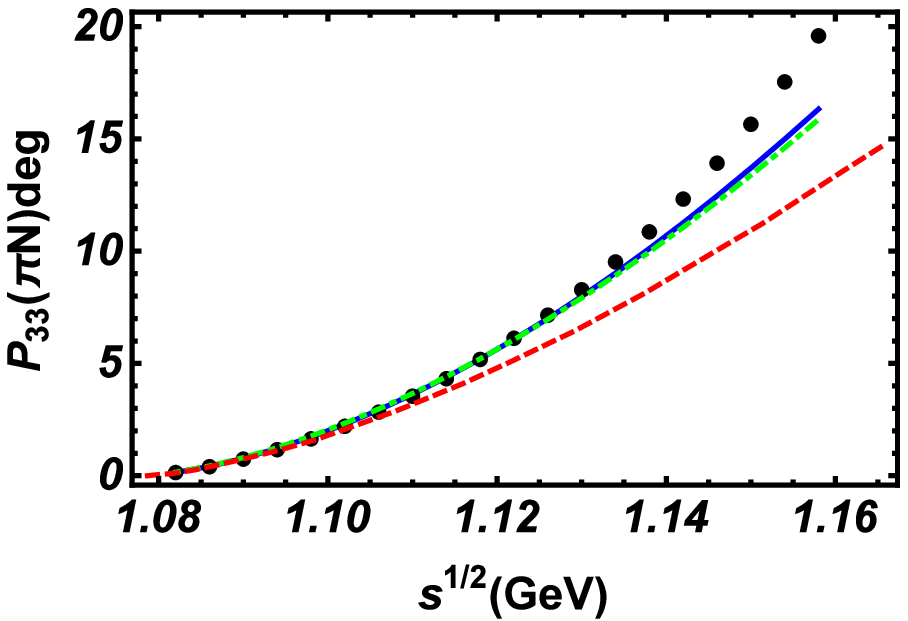}} \\
\end{tabular}
\caption{Pion-nucleon phase shifts. The blue lines denote our results and the dots the WI08 solutions. For the sake of comparison, we show as well
the EOMS SU(2) results~\cite{Chen:2012nx} (green dot-dashed lines) and the HB SU(3) results~\cite{Huang:2017bmx} (red dashed lines).}\label{fig:pin}
\end{figure}

\begin{figure}
\centering
\begin{tabular}{ccc}
{\includegraphics[width=0.25\textwidth]{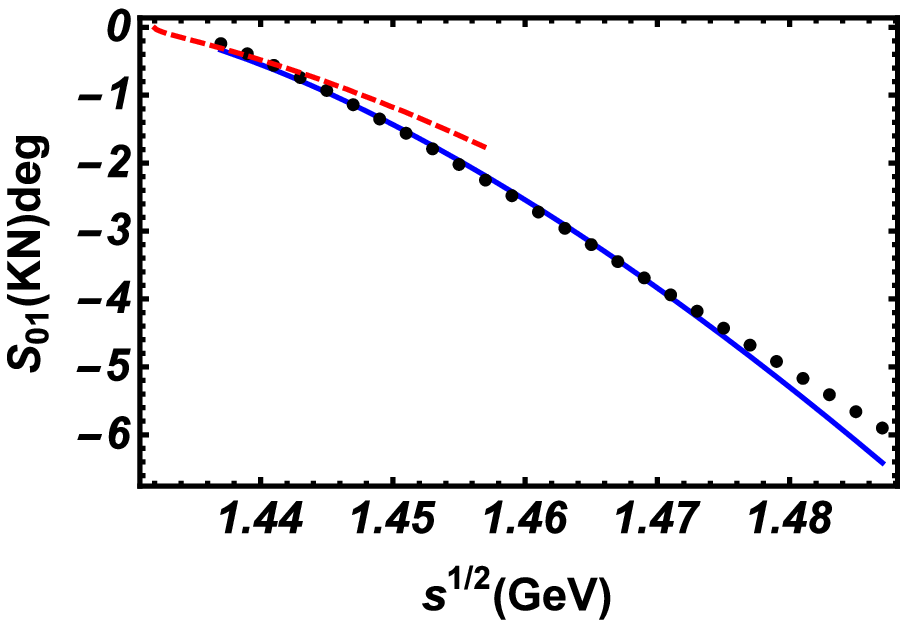}}&{\includegraphics[width=0.25\textwidth]{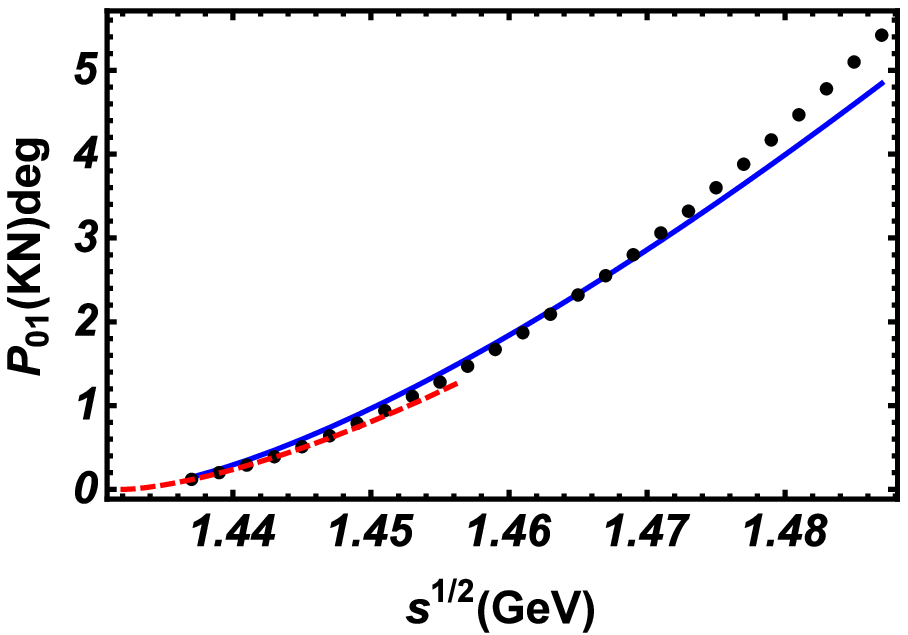}}& {\includegraphics[width=0.25\textwidth]{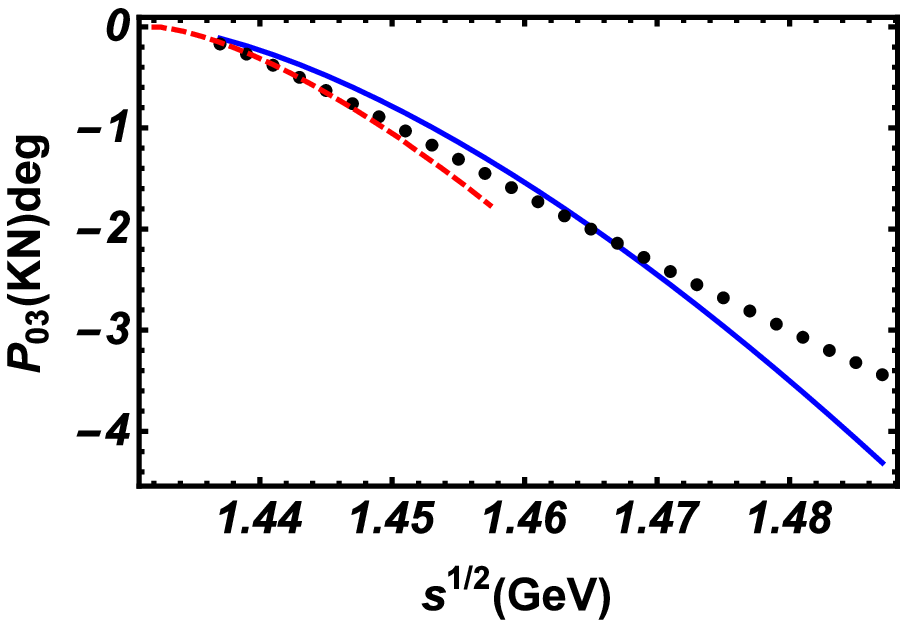}} \\
{\includegraphics[width=0.25\textwidth]{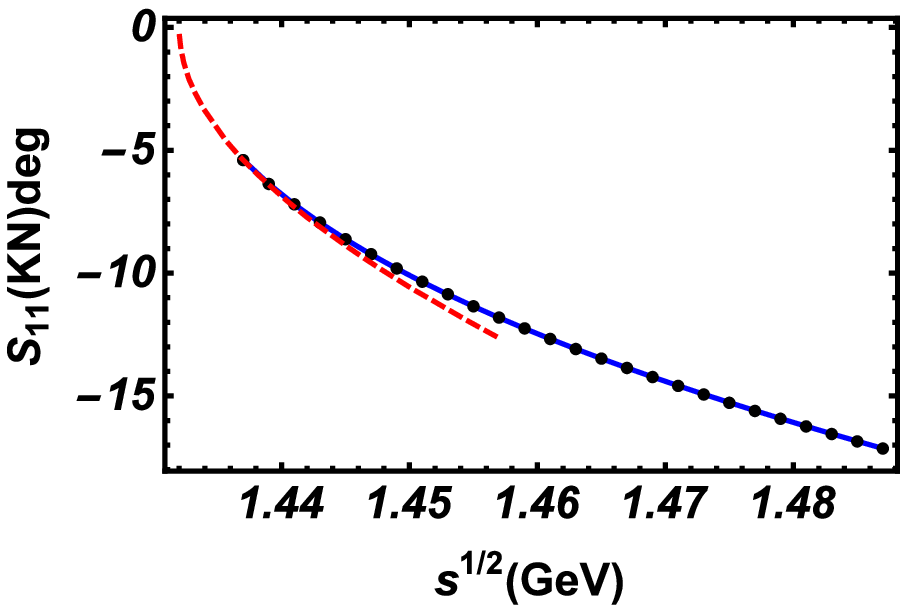}}&{\includegraphics[width=0.25\textwidth]{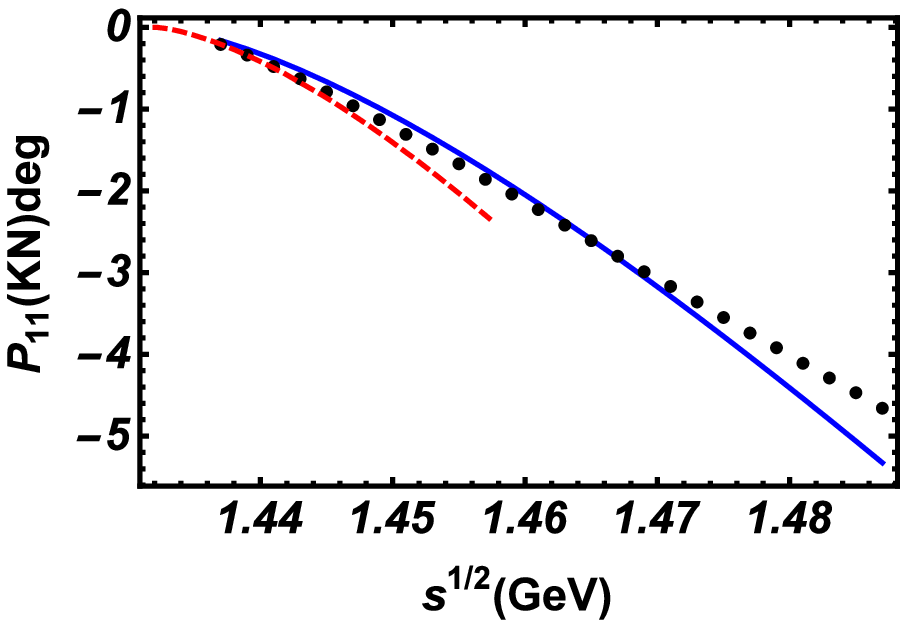}}& {\includegraphics[width=0.25\textwidth]{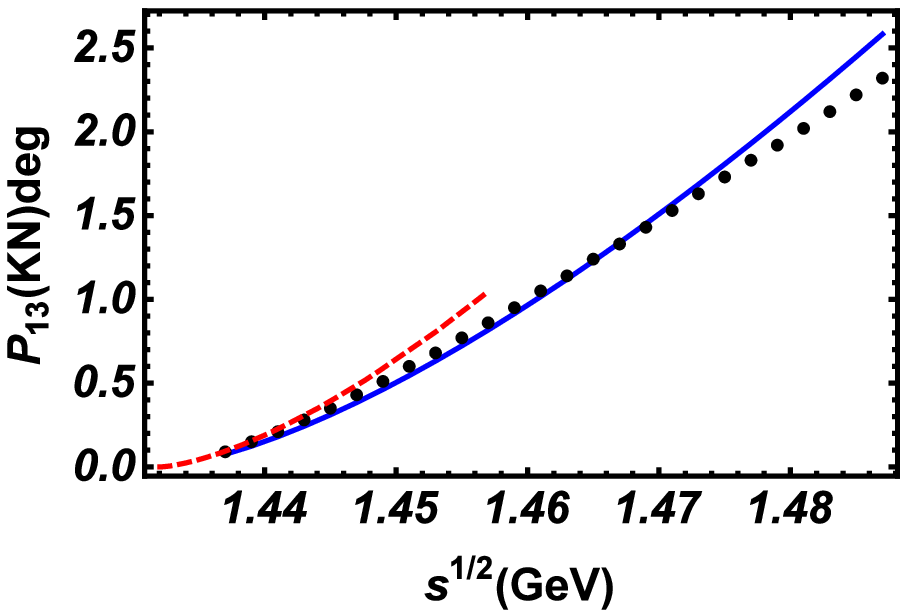}} \\
\end{tabular}
\caption{$I=0$ (upper panel) and $I=1$ (lower panel) $KN$ phase shifts. The blue lines represent our results while the red dashed lines denote those of the  HB ChPT~\cite{Huang:2017bmx}. The dots donate the WI08 solutions.}\label{fig:kn}
\end{figure}

With the amplitudes properly renormalized, we determined the LECs by fitting to the partial wave phase shifts from the analysis of WI08~\cite{Arndt:2006bf} for $\pi N$. And correspondingly, the phase-shift analysis of the SP92 solution~\cite{Hyslop:1992cs} are used for $K N$.
The fitting results are shown in Fig.~\ref{fig:pin} and Fig.~\ref{fig:kn}. The corresponding fit results are compared with the empirical data in both figures.
For the sake of comparison, we show as well the $\mathcal{O}(p^3)$ results of the SU(3) heavy baryon(HB)~\cite{Huang:2017bmx} and the SU(2) EOMS BChPT~\cite{Chen:2012nx}.

Clearly,  the EOMS results can describe the
phase shifts quite well.
While the data are only fitted up to $\sqrt{s}=1.13$GeV, the phase shifts are described very well even up to $\sqrt{s}=1.16$GeV. Besides, our calculation in SU(3) shows a compatible description compared to that in SU(2), which implies that the inclusion of strangeness has small effects on the fitting results.
For the $KN$ scattering, a quite good description of the phase shifts can already be achieved at NLO. We found that in the
EOMS ChPT, the phase shifts alone cannot uniquely fix the eight LECs up to $\mathcal{O}(p^3)$. For the sake of comparison, we show as well the HB results of Refs.~\cite{Huang:2017bmx}. It is clear that the EOMS descriptions are slightly better that the HB results when extended to higher energy region.

\section{Summary}
In this talk, we present our results on meson-baryon scattering up to next-to-next-to-leading order in the framework of covariant ChPT with the EOMS scheme. We determined the LECs by fitting to the partial wave phase shifts from GWU group. We achieved a pretty good description for $\pi N$ and $KN$ channels simulatively. Compared with those in HB scheme, our results in $\pi N$ channels are much better, while in $KN$ channels, the improvements are not significant as in $\pi N$ cases.

\section{Acknowledge}
This work is partly supported by the National Natural Science Foundation of China under Grants No.11522539, 11735003
and the fundamental Research Funds for the Central Universities.

%

\end{document}